\documentclass[aps,prl,preprint,superscriptaddress]{revtex4}

\newcommand{\hbn}{{\em h}BN}
\newcommand{\cmms}[1]{}

\usepackage{graphicx}
\usepackage{psfrag}
\begin{document}

\title{
Anisotropic excitonic effects in the energy loss function 
of hexagonal boron nitride 
}

\author{S.~Galambosi}
\affiliation{Department of Physics, POB 64, FI-00014 University of
  Helsinki, Finland }
\author{L.~Wirtz}
\affiliation{IEMN, CNRS UMR 8520, Dept. ISEN, B.P. 60069, 59652 Villeneuve d'Ascq, France}
\author{J.~A. Soininen}
\affiliation{Department of Physics, POB 64, FI-00014 University of Helsinki, Finland }
\author{J.~Serrano}
\affiliation{ICREA--Departamento de Fisica Aplicada, Universitat
  Politecnica de Catalunya, EPSC, Av. Esteve Terradas 15, 08860
  Castelldefels, Spain}
\author{A.~Marini}
\affiliation{CNISM, Dipartimento di Fisica, Universit\`a di Roma “Tor Vergata”, 
            via della Ricerca Scientifica, 00133 Roma, Italy}
\author{K.~Watanabe}
\affiliation{National Institute for Materials Science, 1-1 Namiki, Tsukuba, 305-0044, Japan}
\author{T.~Taniguchi}
\affiliation{National Institute for Materials Science, 1-1 Namiki, Tsukuba,
305-0044, Japan}
\author{S.~Huotari}
\affiliation{European Synchrotron Radiation Facility, BP 220, F-38043 Grenoble cedex, France}
\author{A.~Rubio}
\affiliation{Nano-Bio Spectroscopy Group, ETSF Scientific Development Centre, 
Universidad del Pa\'\i s Vasco, CFM-CSIC-UPV/EHU-MPC and DIPC, Av. Tolosa 72, E-20018 San Sebastián, Spain}
\author{K.~H\"{a}m\"{a}l\"{a}inen}
\affiliation{Department of Physics, POB 64, FI-00014 University of Helsinki, Finland }

\date{\today}

\begin{abstract}
We demonstrate that the valence energy-loss function of hexagonal boron nitride (\hbn)
displays a strong anisotropy in shape, excitation energy and dispersion for momentum 
transfer {\bf q} 
parallel or perpendicular to the \hbn\ layers. 
This is manifested by e.g. an energy shift of 0.7~eV that cannot be captured by single-particle approaches and is a demonstration of a 
strong anisotropy in the two-body electron-hole interaction.
Furthermore, for in-plane directions of {\bf q} we observe a splitting of the $\pi$-plasmon
in the $\Gamma$M direction that is absent in the
$\Gamma$K direction and this can be traced back to band-structure effects.
\end{abstract}

\maketitle

Layered hexagonal boron nitride (\hbn) is the III-V compound counterpart of
graphite. The recent interest in \hbn\ is partly due to Watanabe et
al.~\cite{Watanabe2004} who showed that \hbn\ exhibits the potential for lasing
at high energies (5.8~eV). This makes \hbn\ an attractive candidate for
optoelectronic applications in the ultraviolet energy range.   Furthermore, a
single layer of \hbn\ is isoelectronic to graphene and can be considered the
precursor material for boron nitride nanotubes.  Despite the relatively simple
crystal structure, \hbn\ appears to be a challenge to both experiments and
theory.  The band-structure of \hbn\ displays an indirect
gap~\cite{Blase1995,Arnaud2006} and the optical absorption spectrum is
dominated by correlation effects leading to a strong Frenkel-type excitonic
peak at 5.8 eV~\cite{Arnaud2006,Wirtz2006,Marini2008} in agreement with the
experimental findings~\cite{Watanabe2004}.  The origin of the fine structure of
the absorption and luminescence spectra around this peak, however, is still
under debate~\cite{Wirtz2008,Arnaud2008,Watanabe2009}.  Recent luminescence
experiments point towards the role of
defects~\cite{Jaffrennou2007,Watanabe2006,Museur2008} in agreement with the
theoretical suggestion in ref.~\onlinecite{Wirtz2008}.

A major source of experimental difficulties lies in the 
growth of large defect-free \hbn\ crystals. 
This is especially important for the electronic 
properties since \hbn\ stacking order can have a substantial effect on the 
band structure~\cite{Liu2003, Ooi2006}. In the case of cubic boron nitride, the
discrepancies between experimental and theoretical results can often be traced
back to the use of low quality samples~\cite{Satta2004}.

In order to study the dynamics (e.g.~possible dispersion) of the valence
excitations one needs to use inelastic scattering probes such as electron
energy loss spectroscopy (EELS) or non-resonant inelastic x-ray scattering
(NRIXS).  NRIXS has the advantage to study  the dispersion of low lying
excitations beyond the first Brillouin zone (1st BZ). This has been used
numerous times in the past (see e.g.~\cite{Caliebe2000,Fuentes2003}) and gives
new important information on the properties of excitations at sub-unit-cell
length scales~\cite{Abbamonte2008}. When studying longitudinal excitations one
often finds non-parabolic dispersions~\cite{Soininen2000} and periodicity for
low energy plasmons~\cite{Galambosi2005,Cai2006}.  The valence bands and the
lowest conduction bands of \hbn\ can be understood as combinations of the
$\sigma$ and $\pi$ states of the hexagonal boron nitride sheet.  At low
momentum transfer the low-energy-transfer structures of the loss function can
be divided into transitions between states with same parity
($\sigma$-$\sigma*$, $\pi$-$\pi*$) when the momentum transfer is in the plane
and different parity ($\pi$-$\sigma*$, $\sigma$-$\pi*$) for momentum transfer
along the c-axis~\cite{Tarrio1989}.  The strong anisotropy in the electronic
response can also be seen in the difference for the dielectric constants
$\varepsilon_\infty$ parallel
and perpendicular to the planes (4.40 and 2.53, respectively~\cite{Arnaud2006}).

In this work, we study the loss function of \hbn\ using NRIXS. We show detailed
experimental results of the dispersion of various features and plasmons for the
momentum transfer {\bf q} in different crystallographic directions within and
beyond the 1st BZ.  The loss features show a strong directional dependence not
only in the comparison in-plane/out-of-plane but also for the different
directions within the plane. The origin of the different spectral features and
their direction and momentum dependence are analyzed by {\it ab-initio}
calculations at the level of the random-phase approximation (RPA) and with
methods of many-body perturbation theory.  In order to exclude a perturbation
of our results by the low crystal quality, we have characterized the sample in
detail.

The sample was a colorless and transparent hexagonal single crystal \hbn\
platelet about 700~$\mu$m wide and 70~$\mu$m thick. The $\Gamma$A direction was
found to be very nearly parallel to the  normal of the platelet.  We used
single-crystal X-ray diffraction to verify that the sample exhibited an AB-type
stacking, thus ruling out other stable or metastable stacking
sequences~\cite{Liu2003,Ooi2006}.

The NRIXS spectra were measured on ID16 at the European Synchrotron Radiation
Facility, Grenoble, France. The experiment was carried out using the
eV-resolution spectrometer~\cite{Huotari2005,Verbeni2009}.  An energy
resolution of 0.6~eV was determined from the FWHM of the quasi-elastic line.
The measurements were performed using the inverse scan
technique~\cite{Hamalainen1996}.  The momentum transfer resolution was  $\Delta
q / q \approx 0.17$ near the K point.

Theoretical NRIXS spectra have been calculated at the level of the random-phase
approximation (RPA) starting with the Kohn-Sham DFT wave-functions in the
local-density approximation (LDA)~\cite{calc_details}.  The dielectric function
is obtained through $\epsilon_{{\bf G},{\bf G'}}(\tilde{\bf q},\omega) = 1 -
v(\tilde{\bf q}+{\bf G}) \chi^0_{{\bf G},{\bf G'}}(\tilde{\bf q},\omega)$,
where $v(\tilde{\bf q}) = 4\pi/|\tilde{\bf q}|^2$ is the Coulomb potential in
reciprocal space and $\chi^0_{{\bf G},{\bf G'}}(\tilde{\bf q}, \omega)$ is the
independent particle polarizability which is obtained from a sum over
transitions from occupied to unoccupied bands~\cite{Adler1962}. ${\bf G},{\bf
G'}$ denote reciprocal lattice vectors, and $\tilde{\bf q}$ is restricted to
the
1st BZ.

The loss function $\sigma({\bf q},\omega)$ is then obtained from the inverse of
the dielectric matrix  \begin{equation} \sigma(\tilde{\bf q}+{\bf G},\omega) =
-\mbox{Im}(\epsilon^{-1}_{{\bf G},{\bf G}}(\tilde{\bf q},\omega)),
\label{lossfunc} \end{equation} where $\mathbf{q}=\tilde{\mathbf
q}+\mathbf{G}$.
Crystal local-field effects
are automatically included by taking into account the off-diagonal elements of
$\epsilon_{{\bf G},{\bf G'}}$ in the matrix inversion~\cite{calc_details2}.  In
order to compare with the experimental loss function, we have used a broadening
$\eta = 0.4$~eV. In silicon the use of the time-dependent LDA kernel leads to a
considerable enhancement in the description of short-range exchange-correlation
effects~\cite{Weissker2006} and to a better agreement with the experiments.  We
have checked that the use of time-dependent LDA does not give a substantial
improvement in the case of \hbn\ .

\begin{figure*} \begin{center}
\includegraphics[width=\linewidth]{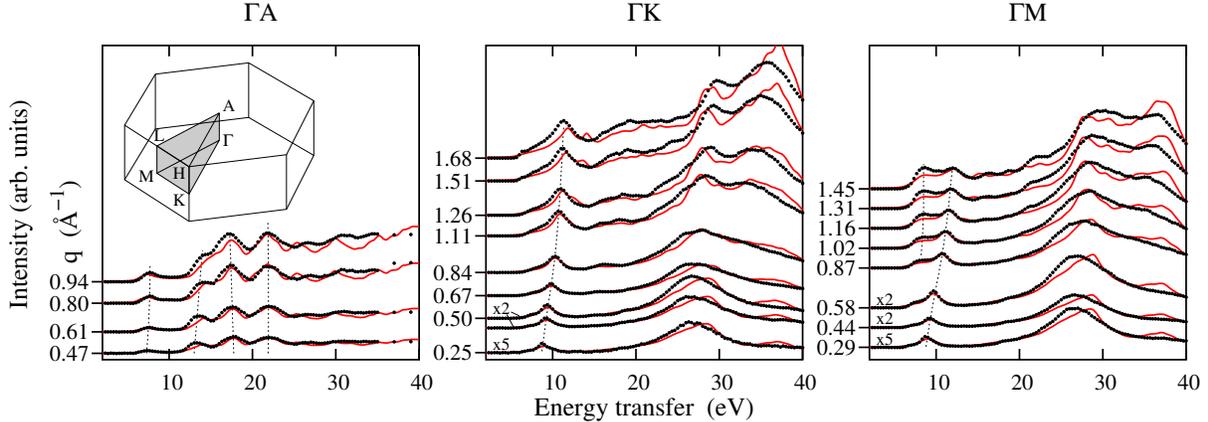}
\caption{ (Color online)
The experimental (black dots) and theoretical (red lines) NRIXS spectra in the
1st BZ along three crystallographic directions. The $q^2$-weighed spectra are
vertically displaced proportionally to the value of momentum transfer, which is
indicated on the vertical axis.  The dashed lines are guides for the eye for
features discussed in the text. The inset shows the 1st BZ of \hbn\ with the
irreducible part shaded.}
\label{1stBZ}
\end{center}
\end{figure*}
\begin{figure}
\includegraphics[width=.9\linewidth]{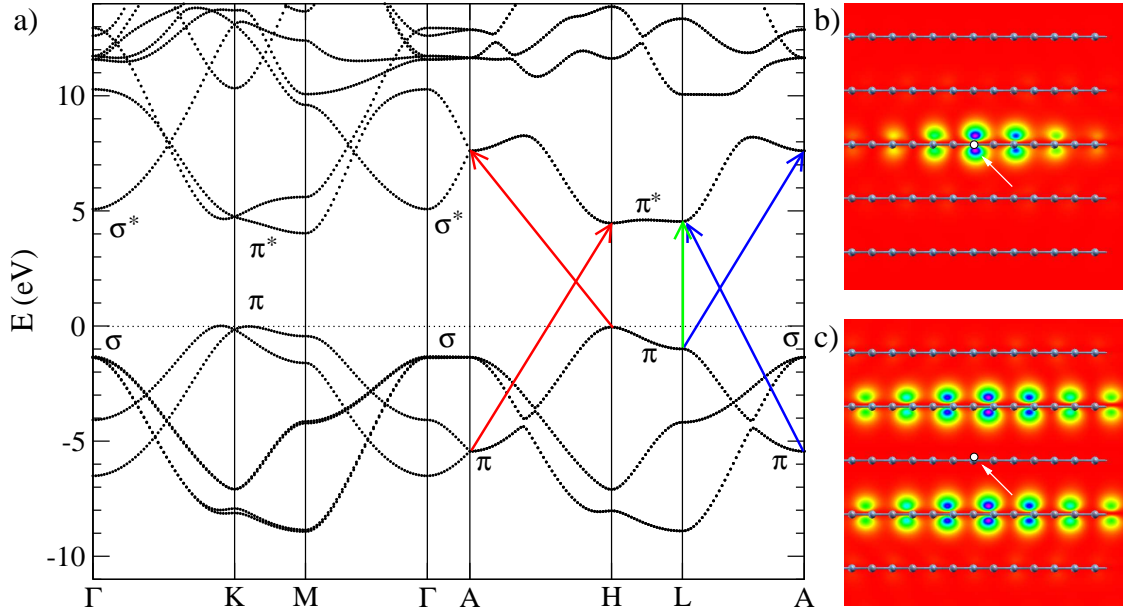}
\caption{(Color online)
a) DFT-LDA band-structure of \hbn.
On the right the 2D projections of the electron probability density
$|\Psi^{\lambda}(r_h,r_e)|^2$ for the lowest bright exciton in the optical
absorption spectrum are shown with ${\bf q} \rightarrow 0$ b) parallel and c)
perpendicular to the \hbn\ planes.  The hole position $r_h$ (marked by white
circle and arrow) is fixed 0.4 a.u.\ above a nitrogen atom. Balls and sticks
indicate the atomic layers.  Calculations of the excitonic wave functions have
been performed using BSE.  } 
\label{splitfig}
\end{figure}

 
The
experimental and theoretical NRIXS spectra along the three crystallographic
directions within the first are depicted in Fig.~\ref{1stBZ}. The experimental
spectra were normalized to have the same area as the calculated spectra in the
energy range shown in the figure.  Overall the calculations reproduce the
experimental spectra well.  The {\em ab-initio} calculations match the
experimental feature positions, their relative weights as well as the momentum
transfer dependence.  From the figure it is evident that the spectral features
along the $\Gamma$A direction do not exhibit any significant dispersion. The
high anisotropy of \hbn\ is clearly reflected in the differences between the
spectra in the hexagonal plane ($\Gamma$K, $\Gamma$M) and the spectrum
perpendicular to it ($\Gamma$A). For $q\lesssim 0.6$\AA$^{-1}$, the spectra
along $\Gamma$K and $\Gamma$M are nearly identical. However, as the value of
$q$ is increased an anisotropy also within the hexagonal plane is clearly
observed.  In the high-energy range this anisotropy shows up mainly as a
different rate of relative spectral weight increase for the feature between
35~eV-40~eV.    The in-plane anisotropy is most evident in the behavior of the
$\pi$ plasmon.  Its energy disperses from 9~eV at small values of momentum
transfer to about 12~eV when $q$ is near the boundary of the 1st BZ. Along
$\Gamma$M an additional peak around 8~eV develops for $q>0.87~$\AA$^{-1}$ while
in the  $\Gamma$K direction only a weak shoulder is detected.  The double-peak
structure along $\Gamma$M is also visible in $\mbox{Im}(\epsilon({\bf q}))$.
This indicates that the directional anisotropy can be interpreted in terms of
interband transitions.  Fig.~\ref{splitfig} (a) shows the band structure of
\hbn.  For ${\bf q}=\Gamma K$, the transitions that dominate the plasmon at 12
eV are the ones from the $\pi$ band at A/H to the $\pi^*$ band at H/A,
respectively (red arrows). For ${\bf q}=\Gamma M$, the dominant transitions are
the ones from the $\pi$ band at A/L to the $\pi^*$ band at L/A (blue arrows).
The observed 8~eV transition at ${\bf q}=\Gamma M$ originates from a
$\pi-\pi^*$ interband transition from L to a neighboring high symmetry point
L'.  This is marked by the vertical green arrow in Fig.~\ref{splitfig} (a) even
though it is not a vertical transition but one with with a momentum difference
of ${\bf q} = \Gamma M$.  At ${\bf q} = \Gamma K$ the plasmon peak around 8~eV
is missing because the joint-density of states displays only a minor peak
there.

The $\Gamma$M and $\Gamma$K  theoretical spectra have been blueshifted by
0.8~eV whereas for the $\Gamma$A spectra a blueshift of 1.5~eV was used. These
shifts reflect the well-known fact that the DFT band-structure usually
underestimates the transition energies between occupied and unoccupied bands.
For \hbn\ it was shown~\cite{Blase1995,Arnaud2006,Wirtz2008} that
electron-electron correlation (calculated on the level of the GW-approximation)
increases the transition energies by about 2 eV with respect to DFT-LDA
calculations.  At the same time, the attractive electron-hole interaction
reduces the transition energy such that the difference between experimental and
theoretical NRIXS spectra is less than 2 eV.  We note that the GW-correction
alone cannot explain the observed anisotropic shift since $\pi$ and $\sigma$
bands are renormalized in the same way.  Therefore the experimental NRIXS data
cannot be explained with single particle theories and the explicit inclusion of
electron-hole interaction is necessary.

\begin{figure}
	\begin{center}
		\includegraphics[width=\linewidth]{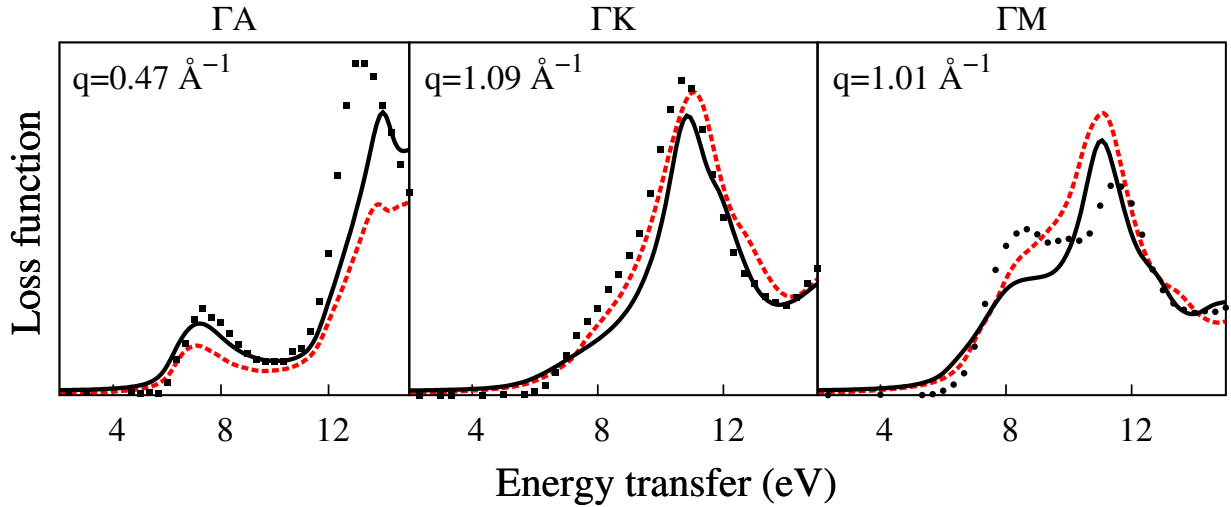}
		\caption{(Color online) The comparison of the current RPA calculations (black solid curve) with a
		BSE calculation (red dashed curve) at selected momentum transfers. The RPA spectra are
		blueshifted as discussed in the text. Experimental data is also shown (black dots).}
		\label{rpa_bse}
	\end{center}
\end{figure}

In order to check if the combined effect of electron-electron and electron-hole
interaction does indeed explain the (anisotropic) shift of the RPA-spectra, we
carried out calculations at selected momentum transfers including excitonic
effects on the level of the Bethe-Salpeter Equation (BSE)~\cite{Onida2002}.  We
used the approach of Refs.~\cite{Benedict1998,Soininen2000} and converged all
the relevant parameters~\cite{BSEpar}. To approximate the quasi-particle band
energies of Ref.~\onlinecite{Wirtz2008}, we used a ``scissor'' of 1.92~eV to
shift the LDA conduction bands energies and a small stretch of 5\% was applied
to the valence bands.  As shown in Fig.~\ref{rpa_bse} the BSE results without
additional energy shifts agree well with the RPA spectra that were blueshifted
to the experimental energy scale. The slight differences between the two
calculations seem to originate mainly from different weights of various
spectral features. Also, it should be noted, that (except for the energy shift)
no extra features seem to arise in the BSE results in comparison with the RPA
calculations.  This confirms that the RPA successfully describes long--range
excitations even in strongly anisotropic and layered materials like \hbn, with
the BSE leading only to minor corrections. Electron-hole effects can alter
extended excitations only when the attraction
is very strong, like in a wide-gap insulators such as LiF~\cite{Onida2002}.

We found that the layered structure of \hbn\ leads to a strongly anisotropic 
electron-hole interaction. This is demonstrated in 
Fig.~\ref{splitfig} b) and c)
where we show the excitonic ``wave functions'' for the lowest visible
excitations in the optical absorption spectrum.
For $q$ parallel to the planes ($q_\parallel$), electron
and hole are localized within one plane  forming
a strongly bound quasi-two dimensional compound~\cite{Arnaud2006,Wirtz2008}. 
For $q$ perpendicular ($q_\perp$), electron and hole are localized on different layers
and form a more weakly bound three-dimensional compound. 
This leads to a larger exciton binding energy for $q_\parallel$ than for
$q_\perp$ and thus explains why the upshift due to the combined e-e and e-h interaction
is lower for $q_\parallel$ than for $q_\perp$.

\begin{figure}
\begin{center}
\includegraphics[width=\linewidth]{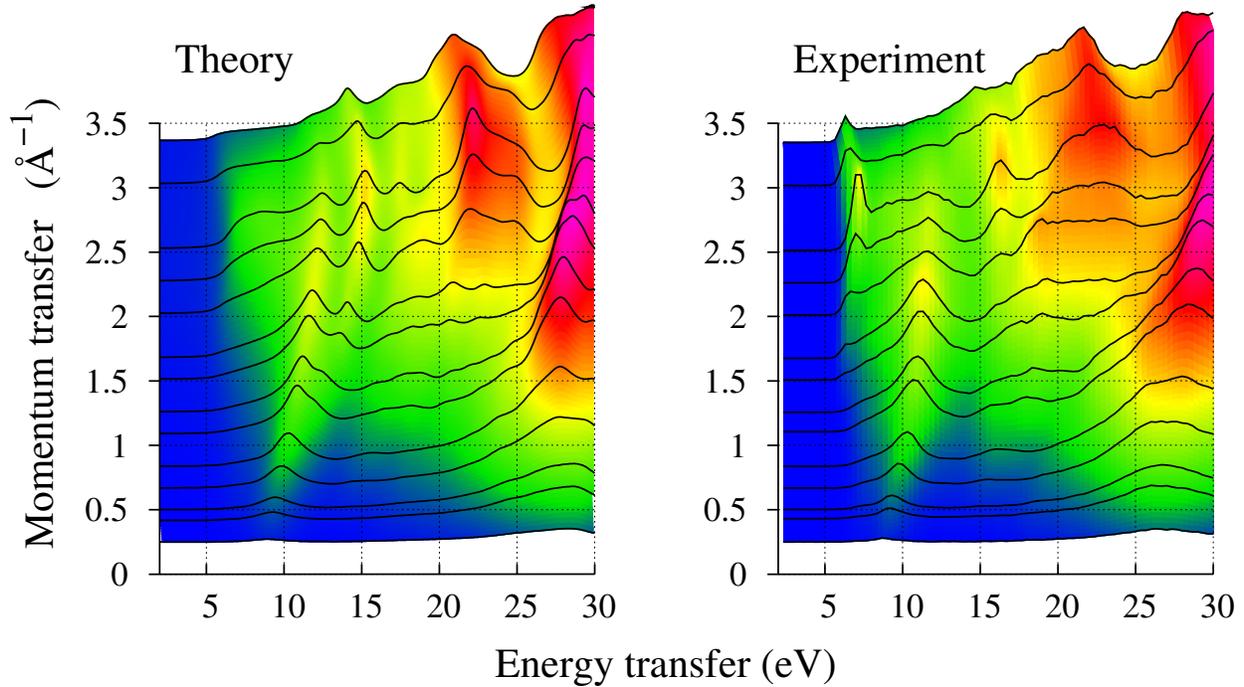}
\caption{(Color online) The theoretical (left) and experimental (right) NRIXS spectra with the
momentum transfer directed along the $\Gamma$K direction using several values
of q.} 
\label{GK}
\end{center}
\end{figure}

The NRIXS spectra with a larger range of momentum transfers directed along
$\Gamma$K are displayed in Fig.~\ref{GK}.  For most parts the calculated
spectra agree well with the experimental ones.  The prominent peak dispersing
between 25-30~eV in the experimental spectra is reproduced very well by the
calculations throughout the measured range of momentum transfers.  At high
momentum transfer ($>2~$\AA$^{-1}$) a new peak showing a small dispersion
appears just above 20~eV. Its appearance and dispersion is reproduced well by
the calculation. In the same momentum transfer range calculation shows a peak
at around 15~eV which in the experiments appears to be less well pronounced and
at a few eVs higher in energy.    Around 7~eV in the high momentum transfer
experimental spectra a sharp peak is
observed, which is not present in the calculated spectra.

In conclusion, we have studied the anisotropy in the valence electron dynamics
of \hbn\ using NRIXS. Besides the in-plane/out-of-plane anisotropy, the
in-plane spectra show a directional dependence at large $q$.  In particular,
the $\pi$-plasmon is split along the high symmetry line $\Gamma$-M but not
along the line $\Gamma$-K.  The anisotropic splitting is an effect of the
anisotropic dispersion of the $\pi$-bands.  We have detected an anisotropy in
the electron-hole interaction for momentum transfer parallel/perpendicular to
the layers.  For $q_\parallel$, the electron and the hole tend to form a
strongly bound quasi-two dimensional compound while for $q_\perp$ electron and
hole tend to be localized on different layers and form a weakly bound
three-dimensional compound. This anisotropy explains the 0.7 eV difference in
the blueshift that must be added to the RPA spectra (which neglect the effect
of electron-correlation) in order to match the experimental data.

This work was supported by the Academy of Finland (1127462/110571), Spanish MEC
(FIS2007-65702-C02-01), "Grupos Consolidados UPV/EHU del Gobierno Vasco"
(IT-319-07), the European Community through  e-I3 ETSF project (GA. 211956).
JAS benefited from the CMS Network RIXS collaboration supported by the U.S.
DoE (grant DE-FG02-08ER46540). AR benefit from fundamental discussions and
collaboration with Prof. T. Pichler.  Computing time was provided by the "Red
Espanola de Supercomputaci\'{o}n" and IDRIS (Proj. No. 091827).


\begin{thebibliography}{34}
\bibitem[{cal({\natexlab{b}})}]{calc_details2}
\bibinfo{note}{The matrix $\epsilon_{\mathbf{G},\mathbf{G}'}$ comprises a total
  of 80 $\mathbf{G}$ vectors in our calculations. The 1st BZ is sampled with a
  60$\times$60$\times$4 $k$-point grid for the calculations with $q$ parallel
  to the hexagonal plane and with a 30$\times$30$\times$12 grid for
  calculations with $q$ perpendicular to it. We have included 8 occupied and 52
  unoccupied bands, up to an energy of 70~eV above the valence band maximum.
  Calculations have been performed with {\tt Yambo} \cite{yambo}.}
\bibitem[{BSE()}]{BSEpar}
\bibinfo{note}{A $k$-space grid of 24$\times$24$\times$12, a $\mathbf{G}$-grid
  of 5$\times$5$\times$5 and 32 unoccupied bands were used in these
  calculations. Finally a broadening of 0.4~eV was applied for the comparison.}
\end{thebibliography}
\end{document}